\documentclass[11pt]{article}
\usepackage{enumitem}
\usepackage{array}
\usepackage{tabu}
\usepackage{mathtools}
\usepackage{amsmath,amssymb,latexsym} 
\setlist{nosep} 

\batchmode

\newtheorem{theorem}{Theorem}
\newtheorem{lemma}{Lemma}

\newtheorem{definition}{Definition}

\def\lor{\mathop{\mbox{\Large $($}}}  
\def\ler{\mathop{\mbox{\Large $)$}}}  

\newcounter{aid}

\begin{document}


\title{\LARGE\bf  Roughly Polynomial Time: A Concept of Tractability Covering All Known Natural NP-complete Problems}

\author{ \vspace*{1mm} Andr\'as  Farag\'o \\
Department of Computer  Science      \\
The  University   of  Texas   at  Dallas\\
   Richardson,  Texas \\
   {\tt farago@utdallas.edu} }

\date{}
\maketitle
\renewcommand{\baselinestretch}{1.05}

\begin{abstract} {\bf\em
We introduce a concept of efficiency for which we can prove that it applies to  all paddable languages,
but still does not conflict with potential worst case intractability.
 Note that the family of paddable languages apparently includes all known natural {\em NP}-complete problems.
 We call our concept {\em Roughly Polynomial Time (RoughP).}  A language $L\subseteq \Sigma^*$, with $|\Sigma|\geq 2$, is in {\em RoughP,} if the following hold: (1) there exists a bijective encoding $\alpha:\Sigma^* \mapsto \Sigma^*$ of strings, such that 
both $\alpha$ and  $\alpha^{-1}$ are computable in polynomial time; 
(2) there is a polynomial time algorithm $\cal A$, which  is an {\em errorless heuristic} for $L$, with exponentially vanishing failure  rate relative to  the $\alpha$-spheres 
$S^{(\alpha)}_n=\{\alpha(x)\,|\;\, |x|=n\}$.
It means, $\cal A$ always correctly decides whether  $x\in L$ or $x\notin L$, whenever it outputs a decision. For some inputs, however, it may not output a decision, rather 
it may  return a special sign,  meaning ``don't know." But the latter  can happen  only  on an exponentially small fraction of each  $\alpha$-sphere $S^{(\alpha)}_n$.
We prove that RoughP contains all paddable languages.
 The result may contribute to the explanation of the often observed gap between practical
algorithm performance  and theoretical worst case analysis for hard problems. Furthermore, the proof also provides a general method to construct the desired encoding and the errorless heuristic. 
Additionally, we also show how to use it
for efficiently generating large, random, guaranteed positive and negative test instances for any paddable language, including all known 
natural {\em NP}-complete problems.
In fact, it appears that every practical decision task (whether in {\em NP} or not) can be represented by paddable languages, and, therefore, our {\bf RoughP} framework applies to all of them. We also explore some connections between {\bf RoughP} and other complexity classes.
}
 \end{abstract}

\thispagestyle{empty}

\newpage
\setcounter{page}{1}

\renewcommand{\baselinestretch}{1}

\section{Introduction and Motivation}

It is a well known phenomenon that algorithms often exhibit better performance 
in practice than what follows from their theoretical analysis. For example,
modern SAT solvers routinely (and successfully!) attack industrial SAT instances with millions of variables, 
despite the conjectured exponential  worst-case running time, as pointed out by Vardi \cite{vardi}. 
This kind of experience, as well as the discontent with the pessimistic view of worst-case complexity, genuinely motivated the search for 
weaker concepts of tractability that could cover {\bf NP}-complete problems, yet avoiding conflict with worst-case hardness. 
This has been a long-standing pursuit, producing a multitude of approaches. 
None of them has led, however, to a reasonable weaker concept of tractability that would be known to cover  
{\em all} {\bf NP}-complete problems, or at least all the intuitively natural ones. In fact, no such broad notion of efficiency has been expected to exist. 

Numerous well known algorithmic concepts pursue, in one way or another, 
the relaxation of the stringent requirement of a worst case 
deterministic polynomial time solution. A few examples: average case analysis; heuristic algorithms (algorithms that may err, but with limited frequency); errorless heuristics (algorithms that never return an incorrect answer, but may fail on some instances); algorithms with 
extra resources (such as randomness or non-uniformity); 
restricting some parameters to constants (fixed parameter tractability); weakening the original question (as in property testing); combining adversarial choices with
random perturbations (as in smoothed analysis); approximations (for optimization versions); and 
a number of others.

While such methods show impressive  success in quite a few cases, none of them offer serious hope to cover {\em all} {\bf NP}-complete problems. 
In fact there are many hardness results, which point in the direction that such a full coverage of {\bf NP} is likely impossible.
Then one can reasonably ask: what if we only want to cover the {\em natural} {\bf NP}-complete problems? After all, they are the ones
that people {\em really} want to solve in practical applications. There are, however, 
two concerns with this:

\smallskip
\begin{description}\itemsep2mm
\item[What is ``natural?"] From the theoretical point of view, there is no definition 
to identify which algorithmic problems are natural. Nonetheless,
this is the smaller issue. After all, for any specific task, there is usually clear consensus whether it is natural or not. For example, 
if a language is constructed by diagonalization, solely for the purpose of exhibiting some complexity phenomenon, then it is viewed artificial.  On the other hand, if a task is motivated by independent interest, or it has already been studied in some different context (such as graph theory, combinatorics, algebra, logic, number theory, programming languages, machine learning, pattern recognition,   etc.),
or it even manifests a practical effort, then its naturalness is rarely debated, if ever.

\item[How to cover at least the naturals?] The bigger problem, however, is this: even if we restrict attention merely to the natural tasks (relying on consensus, rather than definition), still none of the weaker tractability concepts appear  to have the ability to cover all, or even most,  natural  {\bf NP}-complete problems.  
We would like to focus on this issue. 

\end{description}

\smallskip
Let us now take a closer look at {\em heuristic algorithms}, because our approach falls in this class.
Heuristic algorithms come in two primary flavors:

{\bf 1. Algorithms that may err on some inputs.} These algorithms are required to run in
polynomial time, but may return a wrong answer on some inputs.  The key issue here is the error frequency:
on how many instances can the answer be wrong, out of the total of $2^n$ $n$-bit instances? 
({\em Note:} we distinguish this error frequency from the {\em error rate}, by which we mean the 
{\em relative frequency} of errors.) Unfortunately, aiming at low error frequencies runs into conflict with widely accepted hypotheses in complexity theory. For a survey, see
Hemaspaandra and Williams \cite{hema2}. For example, it has been known for a long time that achieving polynomially bounded error frequency is impossible, unless ${\bf P}={\bf NP}$. Subexponentially  bounded error frequency is still known to imply highly unlikely complexity class collapses. 

How about then {\em exponential} error frequency? Note that it can still yield an exponentially low 
error rate. For instance, a $2^{n/2}$ error frequency yields an error rate of 
$2^{n/2}/2^n=2^{-n/2}$. Is that not good enough? The answer is that this task already turns
``too easy:" it allows meaningless trivial heuristics. For example, if we pad an $n$-bit 
input $x$ to $x0^n$, so that it becomes $N=2n$ long, and apply the trivial heuristic 
that accepts {\em all} inputs, then the error rate on the padded language is at most 
$2^{N/2}/2^N= 2^{-N/2}$.  Of course, it does not produce the same error rate when mapped back to
 the {\em original}
problem. But often just the strong asymmetry of yes- or no-instances in the original language 
can already lead to similar trivial cases, without the need for padding. This is quite common, even in natural tasks. For example, regarding the well known {\sc Hamiltonian Circuit} problem in graphs, one can 
prove\footnote{Non-trivially, using methods from random graph theory, see, e.g., Bollobas \cite{bollobas}}
that all but an  exponentially small fraction of $n$-vertex graphs have a Hamiltonian circuit.
Thus, the ``accept everything" trivial heuristic works with exponentially low error rate for this natural problem. Another example is {\sc half-clique}: does the input graph have a clique that contains at least half of the vertices? Here one can prove, using random graph theory again, that the answer is negative for all but an exponentially small fraction of $n$-vertex graphs. Therefore, this {\bf NP}-complete
problem is also solved with exponentially small error rate by a trivial heuristic: reject all instances. Such a trivial heuristic is not meaningful, as it ignores the
very structure we are looking for. 

\smallskip

{\bf 2. Errorless heuristics.} These polynomial time algorithms never output a wrong decision, 
but may fail on some inputs (returning ``don't know"). The error rate is zero, since no error is 
allowed, but there may be a nonzero failure rate. These schemes  have intimate connections to average-case complexity, for a survey see Bogdanov and Trevisan \cite{bogdanov}. Observe that in the errorless
case one cannot simply use a trivial heuristic, capitalizing on the strong asymmetry of yes- or no-instances, as in the above examples. 
It would unavoidably lead to errors, which are not allowed
here at all. That is, the algorithm has to correctly know when to say ``don't know," which may be  rather hard to achieve.

Note that the failure rate can depend on which sets of strings are
used for reference. The traditional way is to count the failures relative to 
all $2^n$ bit strings of length $n$.
Let us call the latter sets the  {\em spheres} of radius $n$, denoted by $S_n$. 
Nothing forces us, however,  to use the $S_n$ as reference sets. If $\alpha$ is a
{\em bijection} on all strings, then we may just as well count the failures on the same sized 
sets $\alpha(S_n)$. If both $\alpha$ and $\alpha^{-1}$ are computable in polynomial time, then 
we call it a {\em $p$-isomorphic encoding}. Observe that 
such a transformation cannot hide much complexity, and it preserves the sphere sizes.
But it may still alter the failure rate, because 
$|\alpha(S_n)|=|S_n|$ does not imply that  the two sets have the same number  of
``don't know"-instances of the errorless heuristic, even though the {\em entire}
set of  ``don't know"-instances, of course, remains the same. This regrouping 
of the instances is somewhat reminiscent to what is called redistricting in election systems.

Our approach can be characterized as an errorless heuristic, which achieves exponentially low
failure rate, capitalizing on an appropriate $p$-isomorphic encoding of the input. The main result is  that this can {\em always} be achieved for paddable languages, which is a very large class.

\section{Notations and Definitions}

Let $\Sigma$ be a finite alphabet, with $|\Sigma|=k\geq 2$,  that we fix for the entire paper. Using standard notation,  $\Sigma^*$ denotes the set of all finite strings formed from the elements of $\Sigma$ (also containing the empty string $\lambda$). It will simplify our treatment if we identify the elements of $\Sigma$ with the numbers $0,1,\ldots,k-1$, each viewed as a single symbol, so we adopt this convention. Subsets of $\Sigma^*$ are referred to as languages.

We use the notation $\mathbb{N}=\{0,1,2,\ldots\}$. The length of a string $x$, i.e., the number of symbols  in $x$, is denoted by 
$|x|$. The length of the empty string is 0. 
If a string $x$ is 
of the form $x=uu$ for some $u\in \Sigma^*$, then $x$ is called {\em symmetric}, otherwise it is {\em asymmetric}. A language $L$ is called {\em trivial} if $L=\emptyset$ or $L= \Sigma^*$, otherwise it is called {\em nontrivial}.

\begin{definition} \label{LM}
{\bf ($p$-isomorphic encoding)} A function $\alpha:\Sigma^*\mapsto\Sigma^* $ is called a
{\em polynomial time isomorphic ($p$-isomorphic) encoding}, if it is a bijection,
computable in polynomial time, and its inverse is also  computable in polynomial time.
\end{definition}

\begin{definition} {\bf (Ball, sphere)}
For any $n\in \mathbb{N}$, the set $B_n=\{x\in\Sigma^*\,|\,\, |x|\leq n\}$ is called the {\em ball} of radius $n$. 
The set $S_n=\{x\in\Sigma^*\,|\,\, |x|=n\}$ is called the  {\em sphere} of radius $n$. For a $p$-isomorphic encoding $\alpha$,
the sets $B^{(\alpha)}_n=\alpha(B_n)=\{\alpha(x)\,|\, x\in B_n\}$ and $S^{(\alpha)}_n=\alpha(S_n)=\{\alpha(x)\,|\, x\in S_n\}$
are called the $\alpha${\em -ball} and $\alpha${\em -sphere}, respectively.
\end{definition}

\noindent
Now we can define {\bf RoughP}, the family of languages 
that are accepted in roughly polynomial time.

\begin{definition} \label{RoughP} {\bf (RoughP)} 
Let $\Sigma$ be an alphabet with $|\Sigma|\geq 2$, and 
let $L\subseteq \Sigma^*$ be a language. 
We say that $L\in {\bf RoughP}$,  if there exist
a $p$-isomorphic encoding  $\alpha$, and a polynomial time algorithm 
${\cal A}: \Sigma^*\mapsto \{{\rm accept, reject,}\, \bot\}$, such that the 
following hold:
\begin{description}\itemsep2mm


\item[\rm \;\;\;\;\; (i)]
$\cal A$ correctly decides $L$, as an errorless heuristic. That is, it never outputs a wrong decision: 
if $\cal A$ accepts a string $x$, then $x\in L$ always holds, and if  
$\cal A$ rejects $x$, then always $x\notin L$.

\item[\rm \;\;\;\;\; (ii)]
Besides accept/reject, $\cal A$ may output the special sign $\bot$, meaning ``don't know" (failure).
This can occur, however, only for an exponentially vanishing fraction of strings in $S^{(\alpha)}_n$.
That is, there is a constant $c$ with $0\leq c<1$, such that 
for every $n\in \mathbb{N}$ 
\begin{equation} \nonumber \label{RP2}
\frac{|S^{(\alpha)}_n\cap \{x\,|\,{\cal A}(x)=\bot\}|}{|S^{(\alpha)}_n|} \leq c^n.
\end{equation}

\end{description}
\end{definition}
{\bf Remark:} It follows directly from the  definition that ${\bf P}\subseteq {\bf RoughP}$, since for $L\in {\bf P}$ we can always 
choose for $\cal A$  the polynomial time algorithm that decides $L$, and use $\alpha(x)=x$. 

\smallskip

A concept that will be important in our treatment is the {\em paddability} of a language. This notion originally gained significance 
from the role it played in connection with the well known Isomorphism Conjecture of Berman and Hartmanis \cite{berman}. The conjecture states that 
all {\bf NP}-complete languages are polynomial time isomorphic ($p$-isomorphic, for short), see \cite{berman}. (Note that a $p$-isomorphism between languages is not the same as our $p$-isomorphic encoding in Definition~\ref{LM}, because the latter does not depend on a particular language.)

Informally, a language is paddable, if in any instance we can encode arbitrary additional information, without changing the 
membership of the instance in the language. Moreover, both the encoding and unique decoding can be carried out in polynomial time. 
To the authors knowledge, all practical/natural decision tasks (whether in {\bf NP} or not) can be represented by 
paddable languages\footnote{This does not mean that every language that represents a practical problem is necessarily paddable. For example, 
it is known that polynomially sparse (nonempty) languages are not paddable (see, e.g., \cite{du}, Theorem 7.15), yet they may still represent practical problems. 
We only say that, to our knowledge, for any practical/natural problem it is
possible to construct a paddable representation, not excluding that there may be other, non-paddable representations, as well.}.
Among the equivalent formal definitions we use the following:

\begin{definition} \label{pad} {\bf (Paddability)} 
A language $L\subseteq \Sigma^*$ is called {\em paddable}, if there exists a polynomial time computable padding function
\, ${\rm pad}: \Sigma^*\times \Sigma^* \mapsto \Sigma^*$ and a polynomial time computable decoding function \, 
${\rm dec}: \Sigma^* \mapsto \Sigma^*$,
 such that for every $x,y\in \Sigma^*$ the following hold:
\begin{description}
\item[\rm \;\;\;\;\;(i)]  $\;{\rm pad}(x,y)\in L$ if and only if $x\in L$.

\item[\rm \;\;\;\;\;(ii)] ${\rm dec}({\rm pad}(x,y))=y$.

\end{description}

\end{definition}

\section{Main Result: All Paddable Languages are in RoughP}

\begin{theorem}\label{thm1}
Let $\Sigma$ be an alphabet with $|\Sigma|=k\geq 2$, and 
$L\subseteq \Sigma^\ast$ be a paddable language. Then $L\in {\bf RoughP}$.
Furthermore, the constant $c$ in {\rm (ii)} of {\em Definition~\ref{RoughP}}
can be chosen as $c=1/\sqrt{k}\leq 1/\sqrt{2}$.
\end{theorem}

\noindent
{\bf Proof.}  If $L$ is trivial\footnote{Recall that $L$ is called trivial if either $L=\emptyset$ or $L= \Sigma^\ast$.
Observe that a trivial language formally satisfies Definition~\ref{pad}, via 
the functions ${\rm pad}(x,y)=y$ and ${\rm dec}(z)=z$.}
 then 
$L\in {\bf P}\subseteq {\bf RoughP}$, so it is enough to consider a nontrivial $L$. 
For the $k$-element alphabet  w.l.o.g. assume $\Sigma=\{0,1,\ldots,k-1\}$.
For any string $x=x_1\ldots x_n\in \Sigma^*$, define $w(x)=x_1+\ldots+x_n$, which we refer to as 
the {\em weight} of $x$. 

Using the paddable language $L$, we define an auxiliary 
language $H\subseteq \Sigma^*$ by
\begin{equation}\label{H'} 
H=\{xx\;|\; x\in L\}\cup \{x\;|\; w(x)\; \mbox{is odd}\}.
\end{equation}
To show that $H$ has  useful properties, let us also define a  polynomial time computable
 auxiliary function $u: \Sigma^* \mapsto \Sigma^*$. Fix two strings $w_0\notin L$, $w_1\in L$ (they always exist 
 for nontrivial $L$), and define  
$u$ as follows:\begin{eqnarray} \nonumber
u(z)= \left\{
\begin{array}{ll}  
x  &  \;\;\mbox{  if  }\;\;z=xx\; \mbox{for}\; x\in \Sigma^* \\ 
w_0 & \;\; \mbox{ if }\;\; \mbox{$z$ is asymmetric and}\; w(z)\; \mbox{is even}\;  \\ 
w_1 & \;\; \mbox{ if }\;\; w(z)\; \mbox{is odd} 
\end{array} \right.
\end{eqnarray}
Recall that a string $z$ is called symmetric if $z=xx$ for some $x\in \Sigma^*$, otherwise $z$ is asymmetric. Symmetry can be easily checked in polynomial time by comparing the two halves of the string (if it has even length, which is obviously necessary for symmetry).  Now we prove some properties of $H$ that we are going to use in the sequel.
\smallskip
\begin{description} \itemsep2.5mm

\item[\rm (a)] 
$L$ has a $\leq^P_m$ (polynomial time many-one) reduction to $H$.  Observe that $x\in L$ if and only if $xx\in H$. (Note that $w(xx)$ is always even, so $xx\in H$ can only occur through the first set on the right-hand side of (\ref{H'}).) Thus, the reduction can be implemented by the function 
$f: \Sigma^* \mapsto \Sigma^*$ defined by $f(x)=xx$, which is clearly computable in polynomial time.

\item[\rm (b)] 
$H$ has a $\leq^P_m$ reduction to $L$. It can be implemented by the function $g: \Sigma^* \mapsto \Sigma^*$ defined as $g(z)=u(z)$. 
To see that it is indeed a  $\leq^P_m$ reduction, consider first $z\in H$. Then either $z=xx$ with $x\in L$, or $w(z)$ is odd. 
In the first case  $u(z)=x\in L$, in the second case $u(z)=w_1\in L$. Therefore, $z\in H$ implies $u(z)\in L$. 
Consider now $z\notin H$. In this case $w(z)$ must be even. Then there are two possibilities: (1) $z$ is asymmetric. Since $w(z)$ is even, we have $u(z)=w_0\notin L$. (2) $z=xx$ for some $x\in \Sigma^*$, but $x\notin L$. Then $u(z)=x\notin L$, so in either case we obtain 
that $z\notin H$ implies $u(z)\notin L$. Thus, noting the polynomial time computability of $u(z)$, we indeed get 
a  $\leq^P_m$ reduction of $H$ to $L$.

\item[\rm (c)]
$H$ is paddable. Using that $L$ is paddable by assumption, let ${\rm pad}(x,y)$ be a padding function for $L$, with decoding function ${\rm dec}(z)$. 
Then a padding function for $H$ can be defined as 
\begin{equation}\label{pad'}
{\rm pad'}(z,y)={\rm pad}(u(z),y)\,{\rm pad}(u(z),y).
\end{equation}
To see that it satisfies Definition~\ref{pad}, take first $z\in H$. Then there are two possibilities:
\smallskip
\begin{description} 

\item[\;\;\;\;\;\rm $\alpha$)] 
$z=xx$ for some $x\in L$, leading to $u(z)=x$. Then  ${\rm pad}(u(z),y)={\rm pad}(x,y)\in L$, due to $x\in L$, from which  ${\rm pad'}(z,y)={\rm pad}(x,y){\rm pad}(x,y)\in H$ follows. 
\smallskip
\item[\;\;\;\;\;\rm $\beta$)]
$w(z)$ is odd, so $u(z)=w_1$. 
Then ${\rm pad}(u(z),y)={\rm pad}(w_1,y)\in L$, due to $w_1\in L$, resulting in   
${\rm pad'}(z,y)={\rm pad}(w_1,y){\rm pad}(w_1,y)\in H$.
\smallskip

\end{description}

Now take $z\notin H$. Then there are again two possibilities:
\smallskip
\begin{description} 

\item[\;\;\;\;\;\rm $\alpha$)] 
$z=xx$, but $x\notin L$. In this case $u(z)=x$, yielding
${\rm pad'}(z,y)={\rm pad}(x,y){\rm pad}(x,y).$
Since ${\rm pad}(x,y)\notin L$, due to $x\notin L$, and  
$w({\rm pad}(x,y){\rm pad}(x,y))$ is always even, therefore, ${\rm pad'}(z,y)\notin H$.

\smallskip
\item[\;\;\;\;\;\rm $\beta$)]
$z\neq xx$ for any $x$, but 
$w(z)$ is even. Then we get $u(z)=w_0$, which gives 
${\rm pad'}(z,y)={\rm pad}(w_0,y){\rm pad}(w_0,y).$
Since ${\rm pad}(w_0,y)\notin L$, due to $w_0\notin L$, and  
$w({\rm pad}(w_0,y){\rm pad}(w_0,y))$ is always even, therefore, ${\rm pad'}(z,y)\notin H$.
\smallskip
\end{description}

Thus, we indeed have ${\rm pad'}(z,y)\in H$ if and only if $z\in H$. 
To get a decoding function ${\rm dec'}$ for $H$, define 
\begin{equation}\label{dec'}
{\rm dec'}(z)={\rm dec}(u(z)).
\end{equation}
We need to show that ${\rm dec'}({\rm pad'}(v,y))=y$ holds for any $v,y\in \Sigma^*$. Observe that (\ref{pad'}) 
and the definition of $u$ imply
$$u({\rm pad'}(v,y))={\rm pad}(u(v),y).$$
Using this, and (\ref{dec'}), we get
$${\rm dec'}({\rm pad'}(v,y))=
{\rm dec}(\underbrace{u({\rm pad'}(v,y))}_{{\rm pad}(u(v),y)})={\rm dec}({\rm pad}(u(v),y))=y,$$
where the last equality follows from (ii) in Definition~\ref{pad}. 
Thus, the function ${\rm dec'}$ indeed carries out correct decoding for ${\rm pad'}$.

\end{description}

\medskip
\noindent
Now we know that  both $L$ and $H$ are paddable. Furthermore,
we have shown  that they are both $\leq^P_m$ reducible to the other. Therefore, it follows from the well known results of Berman and Hartmanis
\cite{berman} that there is a $p$-isomorphism between $H$ and $L$. That is, there exists a bijection
$\varphi: \Sigma^*\mapsto \Sigma^*$, such that both $\varphi$ and $\varphi^{-1}$ are computable in polynomial time,
and for every $x\in \Sigma^*$ it holds that $x\in L$ if and only if $\varphi(x)\in H$.

Let us define the $p$-isomorphic encoding $\alpha$ by $\alpha(x)=\varphi^{-1}(x)$, and define the algorithm $\cal A$ by 
\begin{eqnarray} \label{algo}
{\cal A}(x)= \left\{
\begin{array}{ll}  
\mbox{accept}  &  \;\; \mbox{ if } w(\varphi(x))\; \mbox{is odd} \\ 
\mbox{reject} & \;\; \mbox{ if } w(\varphi(x))\; \mbox{is even and $\varphi(x)$ is asymmetric}\\
\bot & \;\; \mbox{ if $\varphi(x)$ is symmetric.}
\end{array} \right.
\end{eqnarray}
Next we show that this $\alpha$ and $\cal A$ together satisfy Definition~\ref{RoughP}:
\vspace*{1.5mm}
\begin{itemize}\itemsep2mm

\item The function $\alpha$ is a $p$-isomorphic encoding: it is a bijection, plus both $\alpha$ and $\alpha^{-1}$ 
are computable in polynomial time,  due to the same properties of $\varphi$.

\item The algorithm $\cal A$ runs in polynomial time, as $\varphi$ is computable in polynomial time, likewise the symmetry and the parity of 
the weight of any string can be checked in polynomial time.

\item $\cal A$ is an errorless heuristic for $L$, that is, $\cal A$ correctly decides $L$, whenever ${\cal A}(x)\neq \bot$. Indeed, if $\cal A$ accepts, then $w(\varphi(x))$ is odd. This means,
$\varphi(x)\in H$. Then, due to the properties of $\varphi$, it must hold that $x\in L$. Similarly, if $\cal A$ rejects, then $w(\varphi(x))$ is
even and $\varphi(x)$ is asymmetric. This implies $\varphi(x)\notin H$, yielding $x\notin L$. Thus, 
condition (i) in Definition~\ref{RoughP} is satisfied.

\item Finally, it remains to prove condition (ii) in Definition~\ref{RoughP}. 
Let $F=\{z\,|\,{\cal A}(z)=\bot\}$ be the set where $\cal A$ fails. We need to prove 
that there is a constant $c<1$, with 
$$
\frac{|S^{(\alpha)}_n\cap F|}{|S^{(\alpha)}_n|}\leq c^n.
$$
From (\ref{algo}) we know that ${\cal A}(z)=\bot$ if and only if $\varphi(z)$ is symmetric. 
Let $Y$ be the set of all symmetric strings in $\Sigma^*$, then
$F=\{z\,|\, \varphi(z)\in Y\}$.
Consider now the set $S^{(\alpha)}_n\cap F$. The $\alpha$-sphere $S^{(\alpha)}_n$
contains all strings of the form $\alpha(x)$ with $|x|=n$. Among these, those strings $z$ belong to
$F$, for which $\varphi(z)\in Y$ also holds. Therefore, we can write
$$S^{(\alpha)}_n\cap F =\{z\,|\, z=\alpha(x), |x|=n, \varphi(z)\in Y\}.$$
Observe that if $z=\alpha(x)$, then $\varphi(z)=x$, since $\alpha=\varphi^{-1}$. This gives us
$$S^{(\alpha)}_n\cap F =\{z\,|\, z=\alpha(x), |x|=n, x\in Y\}=
\{\alpha(x)\,|\;\, |x|=n, x\in Y\}.$$
The number of symmetric strings among all $n$-long strings is $|\Sigma|^{n/2}$, if $n$ is even, as 
the first half already determines a symmetric string. If $n$ is odd, then their number is 0.
This yields $|S^{(\alpha)}_n\cap F|\leq |\Sigma|^{n/2}=k^{n/2}$. Taking into account that,  due to 
the bijective property of $\alpha$, we have $|S^{(\alpha)}_n|=|S_n|=|\Sigma|^n=k^n$, the bound 
$$
\frac{|S^{(\alpha)}_n\cap F|}{|S^{(\alpha)}_n|}\leq
\frac{k^{n/2}}{k^n}=\left(\frac{1}{\sqrt{k}}\right)^n
$$
follows. Thus, with the choice of $c=1/\sqrt{k}\leq 1/\sqrt{2}<1$ we can indeed satisfy
condition (ii) in Definition~\ref{RoughP}, completing the proof.\\
\hspace*{10mm}\hfill $\spadesuit$

\end{itemize}

\medskip

{\bf Remark.}
The proof actually shows a way to {\em construct} the $p$-isomorphic encoding $\alpha$, and the 
algorithm $\cal A$. Once the $p$-isomorphism $\varphi$, and its inverse  $\varphi^{-1}$ are available, 
$\alpha$ is expressed as $\alpha=\varphi^{-1}$, and $\cal A$ is given by (\ref{algo}). In order to 
obtain $\varphi$ and  $\varphi^{-1}$, recall that we constructed the $\leq^P_m$ reductions $f,g$ 
between $L$ and $H$, as well as the padding/decoding function pair $({\rm pad'}, {\rm dec'})$ for $H$, 
using the the padding/decoding function pair $({\rm pad}, {\rm dec})$ which is assumed available for $L$.
Having the six polynomial time computable functions $f,g, {\rm pad}, {\rm dec}, {\rm pad'}, {\rm dec'}$, 
we can then obtain the $p$-isomorphism $\varphi$ and its inverse $\varphi^{-1}$ via the method of  
Berman and Hartmanis \cite{berman} (see also the textbook description of Du and Ko \cite{du}, Theorem 7.14).
The construction of the $p$-isomorphism is nontrivial, but can be carried out in polynomial time. 
Note that while the expression (\ref{algo}) for  the algorithm ${\cal A}$ may appear deceptively simple, in fact it can be a 
rather complex polynomial time algorithm, since the function $\varphi$ may be complicated.

\section{RoughP and Other Complexity Classes}

From Theorem~\ref{thm1} we know that all paddable languages belong to {\bf RoughP}, and this includes,
among others, all known intuitively natural
{\bf NP}-complete problems, making {\bf RoughP} fairly large. It is then quite reasonable to ask: could it go as far as
${\bf NP}\subseteq {\bf RoughP}$? Another related question is this: if we cannot prove ${\bf NP}\not\subseteq {\bf RoughP}$ 
then which is the smallest mainstream complexity class that is {\em provably}  not a subset of {\bf RoughP}? In this section we present
some claims about these issues.

\begin{lemma}\label{lemma1}
$\;\;{\bf E}\not\subseteq {\bf RoughP}$, where ${\bf E}=\cup_{c>0}{\rm DTIME}(2^{cn})$.
\end{lemma}
\noindent
{\bf Proof.} 
An infinite and co-infinite language $L$ is called {\bf P}-bi-immune,
 if for every infinite $L_0\in \bf P$ it holds that  $L_0\not\subseteq L$ and 
 $L_0\not\subseteq \overline L$.
It is known that {\bf E} contains {\bf P}-bi-immune languages (see Balc\`azar and Sch\"oning \cite{balcazar}). Pick a
{\bf P}-bi-immune language $L\in \bf E$, and assume $L\in {\bf RoughP}$. Let  $\cal A$ 
be the algorithm from
Definition~\ref{RoughP} for $L$, and let $A$ be the set on which $\cal A$ accepts. Then $A\in \bf P$. Furthermore, since $\cal A$ 
is an errorless heuristic, it never accepts
falsely, so $A\subseteq L$. Similarly, let $B$ be the set where $\cal A$ rejects. Again, $B\in \bf P$, and $B\subseteq \overline L$,
as $\cal A$ never rejects falsely.
Due to the failure rate requirement (ii) in  Definition~\ref{RoughP}, $A\cup B$ must be infinite. Therefore, at least one of $A,B$ is infinite, so either 
$L$ or $\overline L$ has an infinite subset in {\bf P}. Thus, $L$ cannot be {\bf P}-bi-immune, a contradiction, proving the claim. \\
\hspace*{10mm}\hfill $\spadesuit$

\smallskip
Note that if {\bf NP} contains a {\bf P}-bi-immune language (which is not known), then the same proof would yield 
${\bf NP}\not\subseteq  {\bf RoughP}$. There is some evidence which supports that 
{\bf NP} may contain a {\bf P}-bi-immune language. Hemaspaandra and Zimand \cite{hema} prove that relative to a random oracle 
{\bf NP} contains a {\bf P}-bi-immune language, with probability 1. Another evidence comes from the theory of resource bounded measure,  
for a survey see Lutz and Mayordomo \cite{lutz}. In this theory a central conjecture is that the $p$-measure of {\bf NP}, denoted by
$\mu_p({\bf NP})$, is nonzero. Informally, this means that {\bf NP}-languages within {\bf E} do not constitute  a negligible subset. 
The $\mu_p({\bf NP})\neq 0$ conjecture can be viewed as a stronger from of the ${\bf P}\neq{\bf NP}$ conjecture, as $\mu_p({\bf NP})\neq 0$ implies
${\bf P}\neq{\bf NP}$, but the reverse implication is not known. Mayordomo \cite{mayordomo} proves that $\mu_p({\bf NP})\neq 0$ implies 
the existence of a {\bf P}-bi-immune language in {\bf NP}, thus reusing the proof idea of  Lemma~\ref{lemma1} for this case yields that 
$\mu_p({\bf NP})\neq 0$ implies ${\bf NP}\not\subseteq  {\bf RoughP}$.

Further contemplating on the ${\bf NP}\subseteq ?\, {\bf RoughP}$ question, observe that while there are plenty of natural problems that are provably in 
${\bf NP}-{\bf P}$, assuming the set is not empty, the situation 
is different with ${\bf NP}-{\bf RoughP}$. The reason is that any $L\in {\bf NP}-{\bf RoughP}$ must be non-paddable, by Theorem~\ref{thm1},
and, of course, be outside {\bf P}. Such  languages in {\bf NP} are in short supply. In fact, it is not known if ${\bf NP}-{\bf P}$ contains a 
non-paddable language, assuming only ${\bf P}\neq{\bf NP}$. The point is that deciding the ${\bf NP}\subseteq ?\, {\bf RoughP}$ 
question in either direction is 
likely to be hard, because in either case it resolves a long-standing, mainstream complexity class separation.

\begin{lemma}
If ${\bf NP}\not\subseteq {\bf RoughP}$, then ${\bf P}\neq{\bf NP}$. If ${\bf NP}\subseteq {\bf RoughP}$, then 
${\bf NP}\neq{\bf EXP}$, where ${\bf EXP}=\cup_{c>0}{\rm DTIME}(2^{n^c})$.
\end{lemma}
\noindent
{\bf Proof.} The first implication follows from ${\bf P}\subseteq {\bf RoughP}$. The second claim is implied by Lemma~\ref{lemma1},
along with ${\bf E}\subseteq {\bf EXP}$. \\
\hspace*{10mm}\hfill $\spadesuit$

{\em Remark:}
Note that ${\bf NP}\subseteq {\bf RoughP}$ also implies ${\bf NP}\neq {\bf E}$, but that is not an open problem, as
${\bf NP}\neq {\bf E}$ has been known for a long time (see Book \cite{book}). But 
${\bf NP}\subseteq {\bf E}$ is not known, in contrast to ${\bf NP}\subseteq {\bf EXP}$.

\smallskip
Another interesting issue is that, in analogy with {\bf NP}, we can also define a class {\bf RoughNP}. Let us  use the  
notation $\langle x,w\rangle$ to represent any standard pairing function (see, e.g., \cite{du}, p.\ 5).  Here $x$ will be  
the instance, and $w$ will represent a witness.

\begin{definition} \label{RNP} {\bf (RoughNP)}
A language $L$ is in the class {\bf RoughNP}, if there exists a language $L_0\in {\bf RoughP}$ and a polynomial $p(n)$, such that for every 
$x\in \Sigma^*$ the following holds: $x\in L$ if and only if there is a $w\in \Sigma^*$, such that
$|w|\leq p(|x|)$ and  $\langle x,w\rangle\in L_0$.
\end{definition}

The definition directly implies ${\bf NP}\subseteq {\bf RoughNP}$, since, due to 
${\bf P}\subseteq {\bf RoughP}$, we can take an $L_0\in \bf P$ in Definition~\ref{RNP}.
In analogy with  ${\bf P}\neq{\bf NP}$, one may conjecture ${\bf RoughP}\neq {\bf RoughNP}$. 
This conjecture may be supported by the following:
\begin{lemma}
If $\mu_p({\bf NP})\neq 0$, then  ${\bf RoughP}\neq {\bf RoughNP}$.
\end{lemma}
\noindent
{\bf Proof.} As shown in the proof of Lemma~\ref{lemma1}, {\bf RoughP} cannot contain a {\bf P}-bi-immune language.
On the other hand, Mayordomo \cite{mayordomo} proves that $\mu_p({\bf NP})\neq 0$ implies 
the existence of a {\bf P}-bi-immune language in ${\bf NP}$. As ${\bf NP}\subseteq {\bf RoughNP}$, this yields ${\bf RoughP}\neq {\bf RoughNP}$.\\
\hspace*{10mm}\hfill $\spadesuit$ 

There are many  more questions that can be raised in connection with the new classes. We plan to address them in the journal version of the paper.

\section{Positive and Negative Test Instance Generation for Paddable Languages}
In this section we present a specific constructive application of the {\bf RoughP} approach: generating large,
random, guaranteed positive and negative test instances for hard algorithmic problems.

For motivation note that in the development of practical algorithms it  is a fundamental need to find appropriate {\em test
instances} to empirically evaluate the performance and correctness of the algorithm. For comprehensive testing 
it is necessary to have large, random test instances, both positive (yes-instances) and negative (no-instances).
Ad hoc solutions of test instance generation for various specific problems have been known for a long time,
in particular for SAT (for an earlier survey see, e.g., Cook and Mitchell \cite{cook}; for state of the art
practical SAT solvers see the International SAT Competitions web page \cite{sat}). 

Arguably, the simplest test instance generation task is when for a given instance length we want to generate 
just a single,  arbitrary positive instance of that length. It is quite natural to ask: can we carry it out efficiently for problems in {\bf NP}? 
The complexity of this problem was studied by Sanchis and Fulk \cite{sanchis}. Among other concepts, they 
introduce the following definition:

\begin{definition} {\bf (PTC)}
A {\em Polynomial Time Constructor (PTC)} for a language $L$ is a deterministic polynomial time algorithm, which, upon input $1^n$,
outputs a string $x$ with $x\in L$, $|x|=n$,  if such a string exists. If there is no such string, then the algorithm outputs 
a special sign $\bot$.
\end{definition}

Unfortunately, it is unlikely that even this simplest instance generation task can {\em always} 
be carried out for problems in {\bf NP}, as
the following theorem can be extracted from \cite{sanchis}: 

\begin{theorem}\label{PTC} {\rm (Sanchis and Fulk \cite{sanchis})}
Every $L\in \bf NP$ has a {\em PTC}, if and only if 
every $L\in \bf P$ has a {\em PTC}, if and only if 
${\bf E}=\bf NE$.
\end{theorem}
Here 
${\bf NE}=\cup_{c>0}{\rm NTIME}(2^{cn})$. The message of Theorem~\ref{PTC} is that 
unless an unlikely collapse happens, there are languages in {\bf NP}, and also in {\bf P}, for which we cannot perform even
this simplest test instance generation task in deterministic polynomial time. 

On the other hand, for those {\bf NP}-complete problems that are deemed  natural, finding a PTC is often quite easy, sometimes outright trivial. 
For example, consider 
the well known {\sc independent set} problem in graphs. To create just {\em any} graph on $n$ vertices containing an independent set of size 
at least $k$, we could simply take $n$ isolated vertices. Of course, this is not viewed as a reasonable test instance, but technically it satisfies the PTC requirements.  

Note, however, that other variants, still within the {\sc independent set} related problem classes, can be significantly harder. For example, considering the search problem for independent sets, Sanchis and 
Jagota \cite{sanchis2} prove the following. As a notation, let us say that for a real number $p$, an $n$-vertex graph has edge density $p$, 
if it has  $\lfloor p n(n-1)/2\rfloor$ edges.
\begin{theorem} {\rm (Sanchis and 
Jagota \cite{sanchis2})}  For any rational number $q\in (0,1)$, and for any real number $p$, 
with $0<p\leq 1-q^2$, there is an integer $n_0$, such that when $n\geq n_0$, and $qn$ is an integer,
there is always a graph with $n$ vertices, maximum independent set of size exactly $qn$, and edge
density $p$. Furthermore, finding a maximum independent set in these graphs is {\bf NP}-hard.
\end{theorem}
Generating a test instance for this class is much less trivial. It would require creating 
a large graph, precisely  with a given edge density, such that its maximum independent set size is exactly $qn$. Finding negative instances efficiently would also be quite nontrivial.

So far we have considered the generation of single, arbitrary instances. As demonstrated with a simple example, this can lead to degenerated
cases. Therefore, for practical purposes, it is much more desirable to generate large {\em random} instances. 

How hard is random instance generation for {\bf NP} languages?
On the one hand, a random instance also passes for an arbitrary instance, with the additional requirement of the random choice from a complicated set.
Hence, we can  expect it to be at least as hard as the PTC problem, which already implies an unlikely collapse (see Theorem~\ref{PTC}). On the other hand, the random instance generator can use the additional power of randomness, which the PTC cannot   use, being deterministic. 
Therefore, they are not directly comparable. But hardness results are still available for random instance generation.  Watanabe \cite{watanabe} proves such a hardness result for distributional 
{\bf NP} search problems. He considers polynomial time computable distributions over the instances, and the generator is required to output a 
certified positive instance with a probability that is polynomially related to the original probability of the instance.
\begin{theorem} {\rm (Watanabe \cite{watanabe})}
If every distributional {\bf NP} search problem, with a polynomial time computable distribution, has a polynomial-time random test instance generator, then ${\bf RE} = {\bf NE}$.
\end{theorem}
Here {\bf RE} is the exponential time analog of {\bf RP}, with linear exponent. The ${\bf RE} = {\bf NE}$ collapse is slightly weaker than 
the ${\bf E}=\bf NE$ collapse  in Theorem~\ref{PTC}, but it is still deemed unlikely. 

In view of the hardness results, it seems reasonable to somewhat relax the requirements. We are going to present a random test instance generator, both for positive and negative instances, such that it  provably always provides {\em guaranteed} positive and negative random instances for any paddable 
language. Recall that this includes all known natural  {\bf NP}-complete problems. The generated instances are uniformly random, but 
possibly not over all instances of a given  length. To capture their distribution, let us introduce  the following  concept.
\begin{definition}\label{Munif} 
{\bf ($M$-uniform distribution})
Let $M$ be a positive integer. A random variable $\xi$ is called $M$-uniform, if there is a set $S$ with $|S|=M$, such that
for every $x\in S$ it holds that $\Pr(\xi=x)=1/M$. 
\end{definition}
Note that this simply means $\xi$ is uniform over $S$, and does not take any value outside $S$, but $S$ may not be known, apart from its size.
In our application $\xi$ will represent the randomly generated instance, but the 
set $S$ will not be explicitly given. Therefore, we will not be able to claim that we generate a uniform random instance from a simple specific set. 
Rather, we can only say that an $M$-uniform instance is generated, with exponentially large  $M$, but $S$ will not be 
explicitly given, apart from polynomial 
lower and upper bounds on most instance lengths in $S$. This can be viewed as a relaxation of a uniformly random instance from all 
instances of a given length.

Now we can define our test instance generator, which we call {\bf RoughP}-generator, since it is based on 
our concept of roughly polynomial time.

\begin{definition} \label{RPG} {\bf (RoughP-generator)} A  probabilistic polynomial time algorithm 
is called a {\em {\bf RoughP}-generator} for a language $L\subseteq \Sigma^*$, with $|\Sigma|=k\geq 2$, if upon receiving the 
input $(1^n,s)$, where $n\in \mathbb{N}$ and $s\in \{pos, neg\}$, the generator 
always outputs a string $x\in \Sigma^*$ in  polynomial time, with the following properties:

\smallskip
\begin{description}\itemsep2mm

\item[\rm \;\;\; (i)\;\;] If $s=pos$,  then $x\in L$ always holds (positive instance).

\item[\rm \;\;\; (ii)\;] If $s=neg$,  then $x\notin L$ always holds (negative instance).

\item[\rm \;\;\; (iii)] There exist   a polynomial $p(n)\geq n$, depending only on $L$, and 
a constant $c> 1$,   such that 
$$\Pr\lor n\leq |x| \leq p(n)\ler \geq 1-c^{-n}$$
where the probability is meant with respect to the internal random choices of the algorithm.

\item[\rm \;\;\; (iv)] 
There is a constant $a>1$, such that the 
output $x$ is $M$-uniform, with $M\geq a^n$.

\end{description} 

\end{definition}

\begin{theorem}\label{RPgen}
Every paddable language $L\subseteq \Sigma^*=\{0,1,\ldots,k-1\}^*$, $k\geq 2$, has a 
{\bf RoughP}-generator, which can be implemented by the following algorithm:

\smallskip
\noindent Upon receiving input $(1^n,s)$, do

\smallskip
\begin{description}\itemsep2mm

\item[\;\;\; Step 1] 
Compute $m=4\lfloor n/2\rfloor +3$. 

\item[\;\;\; Step 2] 
Draw a uniformly random string $z$ with $|z|=m$,
by drawing symbols $z_1,\ldots,z_m$ independently and uniformly at random from $\Sigma$, and setting $z=z_1\ldots z_m$.

\item[\;\;\; Step 3] 
 Compute $w(z)=z_1+\ldots +z_m$. 
 If $s=pos$ and $w(z)$ is odd, or if  $s=neg$ and $w(z)$ is even,  then go to {\rm Step 5.}
 
 \item[\;\;\; Step 4] 
 Draw a number $\nu\in \{1,\ldots,m\}$ uniformly at random. Replace $z_\nu$ in $z$ by another symbol that is chosen 
independently and uniformly  at random among those symbols that have opposite parity to $z_\nu$.

\item[\;\;\; Step 5] 
Output $x=\varphi(z)$, where $\varphi$ is the same polynomial time computable function that is used in the algorithm {\rm (\ref{algo})}.

\end{description}

\smallskip
\noindent
This algorithm satisfies {\rm Definition~\ref{RPG},} such that the constant in {\rm (iii)} is $c=k\geq 2$, and 
the constant in {\rm (iv)} is $a=k^2\geq 4$.
 
\end{theorem}


\noindent
{\bf Proof:} See appendix A.

\section{Discussion}

Our main result is that every paddable language is in {\bf RoughP}. This means, it can be 
recognized by an efficient algorithm in the relaxed sense we have defined: by an errorless
heuristic with exponentially vanishing failure rate over the $\alpha$-spheres. Note that this does not
conflict with potential worst case intractability. 

How large is the set of paddable languages? Apparently, to the author's best knowledge, it
includes all known intuitively natural {\bf NP}-complete problems. 
But how much farther can it go? Surprisingly, it appears that {\em every practical decision problem,} whether in {\bf NP} or not,
has a paddable representation,
when represented by a formal language.  As noted earlier, we  do not mean that all languages that represent a natural problem are necessarily paddable. For example, 
it is known that polynomially sparse (nonempty) languages are not paddable, and they may also represent 
natural problems. We only say that, to our knowledge, for any practical/natural task it is
{\em possible} to construct a paddable representation, not excluding that there may be other, non-paddable representations, as well. 
Since there is no definition of what constitutes a practical decision problem, we cannot make a formal claim here. But we venture into the following (bold) thesis:
\vspace*{-1mm}
\begin{quote}
{\bf Paddability Thesis:} {\em Every practical decision problem has a representation by 
a paddable formal language.}
\end{quote}
\vspace*{-1mm}
\noindent
In itself, this would not be extremely surprising. However, by our results,
we can go further, and assert a thesis, which already becomes provable, once we 
accept the Paddability Thesis.
\begin{quote}
\vspace*{-1mm}
{\bf RoughP Thesis:} {\em Every practical decision problem has a {\bf RoughP} algorithm. Furthermore, it also has a {\bf RoughP}-generator, which can efficiently generate large, random, guaranteed positive and negative instances. }
\end{quote}
\vspace*{-1mm}
\noindent
This thesis sends the unexpected, but still supportable message that {\em every} practical decision problem is solvable with the sense of efficiency 
that {\bf RoughP} offers.

\medskip\medskip

\noindent
{\bf\Large Appendix A}

\medskip\medskip

\noindent
{\bf Proof of Theorem~\ref{RPgen}.} 
Let us consider again the auxiliary language $H$ that we used in the proof of Theorem~\ref{thm1}:
$$
H=\{xx\;|\; x\in L\}\cup \{x\;|\; w(x)\; \mbox{is odd}\}.
$$
We know from the proof of Theorem~\ref{thm1} that $H$ and $L$ are $\leq^P_m$ equivalent (that is, both are $\leq^P_m$
reducible to the other), and we have also proved that $H$ is paddable. As $L$ is also paddable by assumption, there is a 
$p$-isomorphism $\varphi$ between $H$ and $L$, which we have also used, including the algorithm (\ref{algo}).

Observe that  $m=4\lfloor n/2\rfloor +3$ is always an odd number. Since the generated string $z$ has length $m$,
therefore, it always has the property that $z\neq xx$ for any $x\in \Sigma^*$. Consequently, $z\in H$ if and only if 
$w(z)$ is odd. Consider now the following four cases, depending on the value of $s$ and the parity of $w(z)$.

\noindent
{\bf Case 1:}
$s=pos$ and $w(z)$ is odd. In this case $z\in H$, and $z$ is uniformly random over all strings in $H$ with $|z|=m$. 
Since $\varphi$ is a bijection, therefore, the output $x=\varphi(z)$ is uniformly random over the set
$$S=\{ x\,|\, x=\varphi(z),\, |z|=m,\, w(z) \; \mbox{is odd}\}.$$
Furthermore, as $\varphi$ is a $p$-isomorphism, we get that $S\subseteq L$, so all output instances  
are guaranteed to be positive. Regarding the cardinality of $S$,
observe that among all strings $z$, with $|z|=m$, there are at least $\lfloor k^m/2\rfloor$ strings 
for each of the two possible parity values of $w(z)$. This yields $|S|\geq \lfloor k^m/2\rfloor$. From the definition of $m$ we get 
\begin{eqnarray} \nonumber
m = \left\{
\begin{array}{ll}  
2n+3  &  \;\; \mbox{ if } n\; \mbox{is even} \\ 
2n+1 & \;\; \mbox{ if } n \; \mbox{is odd.} 
\end{array} \right.
\end{eqnarray}
Thus, we obtain $|S|\geq \lfloor k^m/2\rfloor \geq \lfloor k^{2n+1}/2\rfloor \geq 
\lfloor k^{2n+1}/k\rfloor = k^{2n}$, so the random output is an 
$M$-uniform positive instance with $M\geq k^{2n}$. This satisfies requirement (iv) in Definition~\ref{RPG} with $a=k^2\geq 4$. 

Considering requirement (iii) in Definition~\ref{RPG}, first observe that due to the polynomial time computability of 
$\varphi$, the length of $x$ is bounded by some polynomial of $|z|=m$. As $m$ is linearly bounded by $n$, there must exist a polynomial 
$p(n)$ with $|x|\leq p(n)$. Moreover, the polynomial depends only on $\varphi$, and for a fixed $L$ we can also fix the $p$-isomorphism, 
implemented by $\varphi$.

For the lower bound $|x|\geq n$, let us estimate $|L\cap B_{n-1}|$, where $B_{n-1}$ is the ball $B_{n-1}=\{x\,|\; |x|\leq n-1\}$.
We can write
$$|B_{n-1}|=\sum_{i=0}^{n-1} |\Sigma|^i = \sum_{i=0}^{n-1} k^i = \frac{k^{n}-1}{k-1} 
\leq k^n-1,
$$ 
yielding $|L\cap B_{n-1}|\leq k^n-1$. Hence, the bijection $\varphi$ can map at most $k^n-1$ strings $z$, with $|z|=m$, into
strings  $x=\varphi(z)$, with $|x|<n$. On the other hand, we already know $|S|\geq k^{2n}$, 
and that $x$ is uniformly random over $S$. Therefore, we obtain 
$$\Pr(|x|<n) \leq \frac{k^n-1}{k^{2n}} < k^{-n}.$$
Taking into account that $|x|\leq p(n)$ always holds, we get
$$\Pr\lor n\leq |x| \leq p(n)\ler \geq 1-k^{-n}$$
with $k\geq 2$.

\noindent
{\bf  Case 2:}
$s=pos$ and $w(z)$ is even. In this case we flip the parity of a random symbol $z_\nu$ in $z=z_1\ldots z_m$ by replacing $z_\nu$ 
with a uniformly random symbol of opposite parity. Then the parity of $w(z)$ also flips, becoming odd. 
By the symmetry of this operation we get that the new string $z'$ is uniformly distributed over the set
$\{ z'\,|\, w(z') \; \mbox{is odd,}\; |z'|=m \}.$ 
Thus, $z'\in H$, and $z'$ is uniformly random over all strings in $H$ with $|z'|=m$, and therefore,  
we are back in Case 1.

\noindent
{\bf  Case 3:}
$s=neg$ and $w(z)$ is even. Then we can repeat the reasoning of Case 1, just replacing the set $S$ by 
$$S'=\{ x\,|\, x=\varphi(z),\, |z|=m,\, w(z) \; \mbox{is even}\},$$
and $L$ by $\overline{L}$. Then we get an 
$M$-uniform negative instance with $M\geq k^{2n}$, satisfying requirement (iv) in Definition~\ref{RPG}. 
Requirement (iii) is satisfied again with the same argument as in Case 1, with the only change of
using $S'$ and $\overline{L}$ in place of $S$ and $L$.

\noindent
{\bf  Case 4:}
$s=neg$ and $w(z)$ is odd. Then we can re-use the reasoning of Case 2: in Step 4 of the algorithm we flip 
the parity of a random  symbol $z_\nu$ in $z=z_1\ldots z_m$ by replacing $z_\nu$ 
with a uniformly random symbol of opposite parity. Then the parity of $w(z)$ also flips, becoming even now. 
By the symmetry of the operation we get that the new string $z'$ is uniformly distributed over the set
of all even weight strings of length $m$. Therefore, we are in the same situation as in Case 3.

Thus, in all cases we established that the requirements (i),...,(iv)  of a {\bf RoughP}-generator 
(Definition~\ref{RPG}) are satisfied.
The running time of the algorithm depends on the time need to compute $\varphi$, about which we know 
it can be done in polynomial time. The additional side computations are clearly done in polynomial time,
which completes the proof. \\
\hspace*{10mm}\hfill $\spadesuit$

\end{document}